\documentclass[prb,aps,showpacs, longbibliography, twocolumn]{revtex4-1}
\pdfoutput=1
\usepackage[pdftex]{graphicx,color} %for arXiv
\usepackage{dcolumn}
\usepackage{bm}
\usepackage{epstopdf}
\usepackage{color}
\usepackage{amssymb,amsmath,amsthm}
\usepackage{lineno}
\usepackage{subfigure}
\usepackage{bibentry}

\newcommand{\tr}{\mathrm{Tr}}

\newcommand{\rank}{\mathrm{rank}\,}

\newtheorem*{lemma*}{Lemma}

\makeatletter
\AtBeginDocument{\@ifpackageloaded{natbib}{\ifNAT@numbers\if@filesw\immediate\write\@auxout{\string\global\string\NAT@numberstrue}\fi\fi}{}}
\makeatother

\begin{document}

\title{Indistinguishability as non-locality constraint} 
\author{C\'assio Sozinho Amorim}   
\affiliation{Department of Applied Physics, Nagoya University, Nagoya 464-8603, Japan}

%\pagewiselinenumbers
\begin{abstract}
A physical explanation for quantum bounds to nonlocality (Tsirelson's bound)  is a fundamental problem that remains open, and one approach to explaining its origins is the so-called Exclusivity principle, relying on probabilistic assumptions shaping general probabilistic theories through sharp measurements and compatible (non-local) measurements. Information indistinguishability, presented here as indistinguishability of qubits and more general bits, may serve as an answer to the nonlocality conundrum, ultimately placing it as the origin to quantum limits. We connect indistinguishability to the exclusivity principle and show that indistinguishability leads to quantum bounds. With that, we suggest indistinguishability to be as fundamental as non-locality and relativistic causality for nonlocal realism.
\end{abstract}

\maketitle
%\tableofcontents

%%%%%%%  Intended Structure %%%%%%%%%%
%  Introduce the idea of state/information indistinguishability, and link it to symmetry
%  Clarify difference between distinguishable and distinct
%  Show how Popescu's work is not "indistinguishable" and how Tsirelson's bound can be seen
%  Connect with Exclusivity Principle
%  乾坤一擲（けんこんいってき）
%%%%%%%%%%%%%%%%%%%%%%%%%%%

%\section{Intro}  %Comment out later
%[Overall flow, from more specific to more general: (i) (re)define qubits and gbit (general bits) with a physical reference (classical bit) for every  encoding bit. (ii) Trace out physical reference for information (gbit) indistinguishability. (iii) Using universal SD, discuss entanglement on QT. (iv) Redefine sharp measurements on GPT to accommodate new gbit definition. (v) point towards E principle as probabilistic constraint.]

Quantum theory: although arcane lore for the laymen, for physicists it constitutes a solid foundation for investigating microscopic phenomena amidst bewilderment that challenges common sense continuously. The nonlocal correlations underlying the theory, such as those pointed in the EPR paradox \cite{Einstein:1935rr}, are examples of such counterintuitive observations. Nonlocality in quantum theory appears by proofs that forbid the theory to become deterministic through local hidden variables, barring away local realism\footnote{By ``realism'' we assume the possibility of measuring some physical property, local or non-local, unambiguously, that is also confirmable by other measurements. A Popescu-Rohrllich box, which breaks translation of perfect correlations and is discussed later, cannot be said realistic in these terms, for while a set of measurements can be combined to lead to the same parity result, one combination gives an opposite, contradictory result for this same physical property.}, 
i.e., the assumption of hidden variables associated with specific local components, and sustaining entanglement as a peculiar resource \cite{Bell:1964wu,Clauser:1969ny,Mermin:1993di,Peres:1993ti,Horodecki:2009gb}. Hence, to preserve realism it is reasonable to accept that nonlocality is fundamental to quantum mechanics and physics, imbuing the theory with nonlocal realism, also confirmed in experiments \cite{Giustina:2015fc,Shalm:2015cw,Hensen:2015dw}. Nonetheless, only the assumption of nonlocality and relativistic causality as axioms fails to explain quantum limits of nonlocal correlation and begs the question of why quantum theory is not  more nonlocal with even stronger correlations, or what makes it suitable for studying microscopic phenomena \cite{Popescu:1994ku}. 
Explanations have been given based on different approaches, namely information causality, macroscopic locality, and exclusivity principle (E principle) \cite{Pawiowski:2009dt,Navascues:2010fk,Cabello:2015ka}. Notably, E principle gives also a bound for the nonlocality of a $n$-body system independently derived by Collins et al. and Seevinck and Svetlichny \cite{Collins:2002ky,*Seevinck:2002hm}.

Another characteristic common to the microscopic world is the indistinguishability of identical particles --- a property identified since Gibbs paradox in statistical mechanics, where the entropy of an ideal gas of particles does not become extensive if one ignores particle indistinguishability. Such property still finds its place in the non-classical world. Entanglement between identical particles has been investigated by different approaches \cite{LoFranco:2016jk,Sciara:2017cz,Sasaki:2011hk,Eckert:2002ij}. While these approaches discuss the entanglement formation related to indistinguishability of bosons and fermions, some aspects may not be so clear when one considers more general situation involving, for example, anyons or states in general probabilistic theories (GPT). This motivates us to consider indistinguishability not only between identical particles (i.e., bosons and fermions), but information units themselves (i.e., qubits and their possible generalizations).

In this paper, by extending the notion of indistinguishability to information units, namely bits, we show that information indistinguishability may explain Tsirelson's bound \cite{Cirelson:1980fp}, including its generalized form for $n$ qubits \cite{Collins:2002ky,*Seevinck:2002hm} by examining the underlying assumptions applied by Cabello \cite{Cabello:2015ka}. We first revisit quantum theory to redefine indistinguishability and qubits, later using it to define information indistinguishability. Then, we show that it leads to entanglement generation between qubits and that the limit for such entanglement coincides with Tsirelson's bound. Finally, we generalize our quantum bits (qubits) to general bits (gbits) in General Probabilistic Theories (GPTs) following Chiribella and Yuan \cite{Chiribella:2014cy} and provide a connection with E principle.

{\it Qubit distinction and information indistinguishability ---} Starting from quantum theory, let us first clarify the word ``indistinguishability'' and redefine qubits. Since one cannot assume indistinguishability to always hold between two particles, which can even bear different statistics (viz. bosonic/fermionic/anyonic), we consider two qubits that may be completely distinct. Henceforth, the word ``distinct''  shall be used to indicate particles (in first quantization) or modes (in second quantization) that can be unambiguously characterized as different by arbitrary means, like a proton and an electron by their charge and mass, or states in hybrid systems. On the other hand, we resort to the word ``distinguishable'' to imply the possibility of identification of particles or states (see for instance \cite{Ivanovic:1987bh,Dieks:1988iy,Kawakubo:2016jd}). We say two distinct qubits are rendered ``indistinguishable'' when their internal states are mixed (by coupling or scattering), and their information cannot be uniquely tracked. For example, a proton and an electron, whose spins we cannot independently track in the ground state of a hydrogen atom without breaking some symmetry to attach each spin unambiguously to each particle, are distinct but their spins become indistinguishable.

For a separable state, a pair of (distinguishable) qubits is usually represented simply by taking the direct product of their kets, i.e., $|q_1\rangle|q_2\rangle$.
 By this representation, each qubit is identified by the order they appear, which reflects in the result of their tensor product. We hence make this label explicit, rewriting the bases as $|r_1,q_1\rangle|r_2,q_2\rangle$, where $r$ is a label that physically identifies the qubit and provides a reference against which a qubit encoding information $q$ is defined. For example, one could write $|e^-,0\rangle|p,1\rangle$ for a hydrogen atom, or $|\mathrm{NV},0\rangle|\mathrm{SC},0\rangle$ for a NV-center/superconducting qubit hybrid system. The reference $r$ becomes a distinguishing (classical) bit in such cases.

\begin{figure}[tb]
\includegraphics[width=0.5\textwidth]{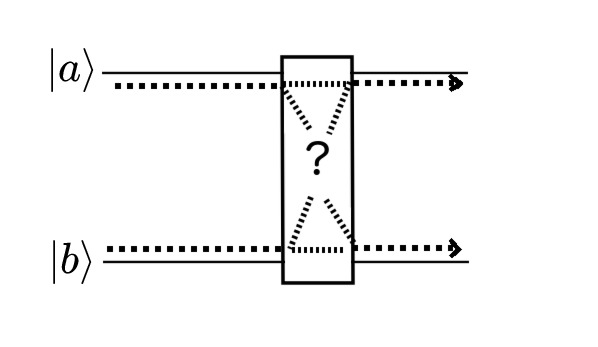}
\caption{Two qubits $|a\rangle$ and $|b\rangle$ (with reference input implicitly identified by their lines) going through an unknown entangling transformation. Given only the output, and the knowledge of the existence of a two-qubit transformation, it is impossible to tell with certainty the path followed by information encoded in qubits $|a\rangle$ or $|b\rangle$ is the control or the target. For instance, for control unitary gates, the target/control identification is inaccessible in the region of the gate if their information is mixed.}
\label{cnot}
\end{figure}

When the relevant reference information in $r$ is inaccessible, we say we have ``information indistinguishability.'' In other words, information indistinguishability is postulated as indistinguishability of qubits alone and becomes physically relevant when an identifying physical background cannot be  associated with a qubit (an internal state). This happens when an entangling gate couples the qubits together (see fig. \ref{cnot}), transforming $|r_1,q_1\rangle|r_2,q_2\rangle$ into $|q\rangle| q'\rangle$. In this scenario, without the clear association of one qubit with one physical holder, there is no longer a clear difference between the two bits, which become essentially indistinguishable. It can also be stated in Lo Franco and Campagno's term of lack of which-way information\cite{LoFranco:2016jk}: a situation in which an internal state cannot be associated with a physical bit in particular. Under this situation, the qubits' encoded information cannot be defined regarding their reference $r$ but only regarding each other, i.e., $q$ becomes $q'$ reference and vice-versa. As so, label $r$ cannot be simply erased but must be eliminated consistently. 

Note that a state may also be  in a superposition or mixture of distinguishable and indistinguishable states, leading to partially (in)distinguishable states \cite{Ivanovic:1987bh,Dieks:1988iy,Kawakubo:2016jd}. A state where each qubit can (not) be perfectly identified is said to be perfectly (in)distinguishable. When qubit bases are perfectly indistinguishable, we expect to have maximum entanglement.

%\begin{figure}[bt]
%\includegraphics[width=0.5\textwidth]{cnot.pdf}
%\vspace{-10 pt}
%\caption{Two qubits $|a\rangle$ and $|b\rangle$ (with reference input implicitly identified by their lines) going through an unknown entangling transformation. Given only the output, and the knowledge of the existence of a two-qubit transformation, it is impossible to tell with certainty the path followed by information encoded in qubits $|a\rangle$ or $|b\rangle$ is the control or the target. For instance, for control unitary gates, the target/control identification is inaccessible in the region of the gate if their information is mixed.}
%\label{cnot}
% Exclude single qubit gates
% Not a two qubit gate, but a control gate. sqrt{SWAP} too?
% But SWAP = CNOT^3, so what's the matter?
%\end{figure}

{\it Excluding physical background ---} 
To ignore the physical background information in $r$, we adapt the procedure in refs. \onlinecite{LoFranco:2016jk, Sciara:2017cz}, introducing a state symmetrization in  partial inner products. The original process from where we depart consists of first defining a symmetric (unnormalized) inner product:
\begin{eqnarray}
\langle \xi|\psi,\phi\rangle = \langle\xi|\psi\rangle|\phi\rangle + \eta\langle\xi|\phi\rangle|\psi\rangle,
\label{sym_inner}
\end{eqnarray}
where $\eta=\pm1$ is a factor accounting for the statistics of the particles. In this inner product, the symmetry imposed is that of particle permutation, according bosonic of fermionic statistics. This product works as a projection of a two-body state of indistinguishable particles onto a single-body state of them. 

We need to change this product to perform two tasks: (a) impose symmetries other than particle permutation, that may account for qubit indistinguishability and (b) impose such symmetry by dropping their distinguishing information (background reference) and establishing the components as mutual references. Hence, one can rewrite eq. (\ref{sym_inner}) as
\begin{eqnarray}
\langle r,r'| r,\phi;r'\psi\rangle &=& \langle r | r,\phi\rangle \langle r'|r',\psi\rangle 
+ \eta\langle r | r,\bar{\phi}\rangle \langle r'|r',\bar{\psi}\rangle,\quad
\end{eqnarray}
where the bar indicates some symmetry transformation gained by redefining references, like time reversal or particle-hole symmetry. 

For our purposes, $|\eta| = 1$ is the only constraint we need, since qubits do not bear any specific statistics by themselves. Assuming coding bases $|q\rangle\in\{|0\rangle,|1\rangle\}$ span $|\phi\rangle$ and $|\psi\rangle$, we will consider symmetries where each bases $|\bar{q}\rangle \equiv |q \oplus 1\rangle$ replaces $|q\rangle$ in the second term of the symmetrization, where ``$\oplus$'' indicates addition modulo 2. Such operation stands for symmetries like time reversal, particle-hole, or other analogous parity preserving symmetries. This form of symmetrization introduces an equivalence between bases of same parity, and is the kind of symmetry we require to impose indistinguishability between them. 

This generalization also makes states $|0, \vec{s}\rangle$ and $|1, \vec{\bar{s}}\rangle$ become indistinguishable, where $\vec{s}$ indicates a general superposition for the second qubit, viz. $|\vec{s}\rangle=\cos\frac{\theta}{2}|0\rangle+e^{i\varphi}\sin\frac{\theta}{2}|1\rangle$, and $\vec{\bar{s}}$ is obtained by exchanging $|0\rangle$ and $|1\rangle$. We therefore realize a state of two qubits where only their relative value has meaning by imposing a symmetry where states of same parity are equivalent.

Once we establish the above base equivalence,  we proceed as follows: 
\renewcommand{\labelenumi}{(\roman{enumi})}
\begin{enumerate}
\item starting from distinct qubits $|r,\phi\rangle|r',\psi\rangle$, the state is expanded in computational bases $|r,0;r',0\rangle, |r,0;r',1\rangle, |r,1;r',0\rangle, |r,1;r',1\rangle$; %why distinct?
\item each basis $|r,q;r',q'\rangle$ is converted into $|q,{q'}\rangle+\eta|\bar{q},\bar{q}'\rangle$, with  $\bar{q}=q\oplus 1$, $|\eta|=1$. This transformation enforces indistinguishability between the bases, with each qubit serving as a mutual reference and only their relative inclination (or parallel/antiparallel bases) remaining.
\end{enumerate}
Step (ii) introduces information indistinguishability in the system and can be recognized to hold equivalence to an entangling (disentangling) gate. 

One may wonder how can the linear combination of two bases account for indistinguishability, which is expected to induce linear dependence of the indistinguishable bases. When related by symmetry, the indistinguishable basis correspond to symmetric images of one another upon a certain symmetry action, and only one of them should make sense, the other becoming a correspondent in a ``virtual'' space, like images in a mirror. On such assumption, one should use the eigenstates of the symmetry supporting indistinguishability (in this case, Bell states) as bases to span  the symmetric space, with states of the same parity related by symmetry becoming linearly dependent.

To clarify the above perspective, we divide the whole original space (without symmetry) into ``symmetric'' space (the relevant half sector after symmetry is imposed) and ``virtual'' space (the remaining half replicated by symmetry). The corresponding ``virtual'' bases (i.e., those corresponding to a basis by symmetry transformation) can be treated as linearly independent on the whole space combining virtual and symmetric space together (see Supplementary information). 
The bases combined in (ii) live in the symmetric and virtual sectors, respectively. Although they do not need to be handled together (one can choose to work with one of them in symmetric space, either $|q,q'\rangle$ or $|\bar{q},\bar{q}'\rangle$), taking the whole space together facilitates to analyze all the effects of the relevant symmetry. In the next sections, we shall show that, starting from a general state in the symmetric space, we obtain entanglement by looking at their Schmidt decomposition (SD), applying the symmetrization discussed here.

{\it Schmidt decomposition for indistinguishable qubits ---} 
It is possible to define an SD following Sciara et al. \cite{Sciara:2017cz} and show that the Schmidt rank obtained is greater than 1 for qubits under indistinguishable condition. Later, we shall see that the upper bound for such entanglement coincides with Tsirelson's bound.

In order to perform an SD, one must look into the density operator $\rho$ of a given state and (i) reduce it to a single particle (qubit) operator. For this, eq. (\ref{sym_inner}) is employed to project the density operator onto single particle basis. Then, (ii) by diagonalizing the reduced density matrix $\rho^{(1)}$, the retrieved eigenvectors can be adequately put together to form Schmidt bases, balanced by the singular values (i.e., the square root of eigenvalues) for each eigenstate. 

For example, consider two bosonic qubits in a state $|\Phi\rangle=|0, \vec{s}\rangle$. The single particle density matrix is obtained by performing a partial trace with inner products defined as eq. (\ref{sym_inner}), leading to
\begin{eqnarray}
\rho^{(1)}=\frac{1}{2N}\left(
\begin{array}{cc}
a & c\\
c^* & b
\end{array}\right),
\end{eqnarray}
with $a=4\cos^2\frac{\theta}{2} + \sin^2\frac{\theta}{2}$, $b= \sin^2\frac{\theta}{2}$,  $c=e^{i\varphi}\sin\theta$, and $N=1 + \cos^2\frac{\theta}{2}$. Schmidt bases can then be derived simply by diagonalizing the reduced density matrix, which in this example gives two eigenvalues
\begin{eqnarray}
\lambda_0=\frac{2}{N}\cos^4\frac{\theta}{4}\quad\mathrm{and}\quad
\lambda_1=\frac{2}{N}\sin^4\frac{\theta}{4}
\end{eqnarray}
that are the square of the singular values giving the weight of the respective eigenstates
\begin{eqnarray}
|\tilde{0}\rangle& = &\cos\frac{\theta}{4}|0\rangle+\sin\frac{\theta}{4}|1\rangle, \nonumber\\
|\tilde{1}\rangle& =& -\sin\frac{\theta}{4}|0\rangle+\cos\frac{\theta}{4}|1\rangle.
\end{eqnarray}
The state can then be written in Schmidt bases as
\begin{eqnarray}
|\Phi\rangle = (\sqrt{\lambda_0}|\tilde{0},\tilde{0}\rangle+\sqrt{\lambda_1}|\tilde{1},\tilde{1}\rangle).\label{bell}
\end{eqnarray}

Notice that before the SD, we have one qubit working as the reference, in this case, set to zero, and one generic superposition for the coding bit. This state is also equivalent, in our set up, to the state $|1,\vec{\bar{s}}\rangle$, and some degree of entanglement is expected. The von Neumann entropy can be calculated as $S = -\sum_i\lambda_i\log\lambda_i$, which in the extreme case of $\theta=\pi$ gives exactly one bit of entanglement entropy. This is expected and can be understood as the product of indistinguishability between bases $|0,1\rangle$ and $|1,0\rangle$. 
The same does not happen to the state $|0,0\rangle$ explicitly for it is mathematically considered distinguishable from $|1,1\rangle$ in this setup. Nevertheless, it is worth noticing that the obtained Schmidt bases for $|0,1\rangle$  in eq. (\ref{bell}) changes into a linear combination of $|+,+\rangle$ and $|-,-\rangle$ states, where $|\pm\rangle=\frac{1}{\sqrt{2}}(|0\rangle \pm |1\rangle)$, which can be associated with a $|0,0\rangle$ and $|1,1\rangle$, respectively. This approaches universalize standard SD and permits to discuss indistinguishable states (see also Appendix \ref{app:SD}).

 {\it Schmidt rank under indistinguishability ---} 
 The non-zero von Neumann entropy is an indicative of entanglement between indistinguishable bases \emph{per se}, but one can go further and look at Schmidt's rank for states perfectly indistinguishable, i.e., occupying only indistinguishable states. 
 
Initially,  let us first define a projector on an indistinguishable subspace of the system, on computational basis, as
\begin{eqnarray}
\Pi = \sum_{ij} |ij\rangle\langle ij|,
\end{eqnarray}
with the summation running over the indistinguishable bases, i.e., those mixed during a two-qubit gate transformation. On the other hand, consider an arbitrary state living in a Schmidt space, spanned by bases calculated according to the recipe described above. We can now make a projector on such space $S$ given by
\begin{eqnarray}
S=\sum_k |\tilde{k}\tilde{k}'\rangle\langle \tilde{k}\tilde{k}'|,
\end{eqnarray}
where the summation runs up to Schmidt's rank and $\tilde{k}$ and $\tilde{k}'$ may be equal or orthogonal depending on the system (e.g., bosonic or fermionic). One can prove that if $||S-\Pi||<1$, the two projectors have the same rank (see Appendix \ref{app:lemma}). This is the case when the state in question lives only in the indistinguishable subspace given by the image of $\Pi$, forcing the Schmidt space to have the same rank as the number of indistinguishable bases within the projector $\Pi$ (see Appendix \ref{app:SR}). The connection between Schmidt rank and $\Pi$ rank tells us that when we have indistinguishable bases in the state space (e.g., $|01\rangle$ and $|10\rangle$), we are expected to be in an entangled set-up between the bases. In other words, the indistinguishability of these bases generates entanglement between them.
%This invites us to extrapolate this result and behold indistinguishability as the generator of any entanglement.

{\it Maximum entanglement for two indistinguishable qubits ---} 
Although Schmidt rank gives us some information about the presence of entanglement, it is not a complete measure of it. However, we may utilize the projector $\Pi$ defined above to reach Tsirelson's bound for the Clauser-Horne-Shimony-Holt (CHSH) correlations \cite{Clauser:1969ny}, i.e.,
\begin{eqnarray}
S_2 = \langle A_1B_2\rangle + \langle A_1B_2\rangle + \langle A_2B_1\rangle - \langle A_2B_2\rangle,
\end{eqnarray}
where $A_m$ and $B_n$ are operators acting on two parts of a system with outcomes $\pm1$ and their indices $m,n\in\{1,2\}$ indicate different bases.

We start by expanding local operators in terms of local POVMs $\mathcal{O}$ as
\begin{eqnarray}
A_m = \sum_i c_i^{(m)}\mathcal{O}_i(r_A),\quad B_n = \sum_i c_i^{(n)}\mathcal{O}_i(r_B),
\label{ops}
\end{eqnarray}
acting on two parties $A$ and $B$, and $i$ identifying orthogonal POVMs spanning $A$ and $B$. Their direct product $A_mB_n$ gives us the total nonlocal operator representing a joint measurement on the bipartite system that will become our entanglement witnesses. We may compute the maximum expectation value for correlations based on these operators for states spanned by indistinguishable bases by taking the inner product with the projectors on the indistinguishable space $\Pi$, i.e.,
\begin{eqnarray}
\langle A_mB_n\rangle &=& \tr(\Pi A_mB_n)\nonumber\\
&=& \tr\left(\Pi \sum_{ij}c_i^{(m)}c_j^{(n)}\mathcal{O}_i(r_A)\mathcal{O}_j(r_B)\right)\nonumber\\
&=&\sum_{ij}c_i^{(m)}c_j^{(n)}\tr\left(\Pi\mathcal{O}_i(r_A)\mathcal{O}_j(r_B)\right)\nonumber\\
&=&\sum_{i\ne j}c_i^{(m)}c_j^{(n)}\tr\left(\Pi\mathcal{O}_i(r_A)\mathcal{O}_j(r_B)\right) \nonumber\\
&+& \sum_{k}c_k^{(m)}c_k^{(n)}\tr\left(\Pi\mathcal{O}_k(r_A)\mathcal{O}_k(r_B)\right).
\label{q_corr}
\end{eqnarray}
In the last equality in eq. (\ref{q_corr}), the first term vanishes, and we obtain the maximum correlation when the trace in it equals unity. Since the coefficients obey normalizing conditions $\sum_i|c_i^{(m)}|^2=\sum_i|c_i^{(n)}|^2=1$, one may check maximum correlations to occur when $A_1$ and $A_2$ have each one (mutually orthogonal) component and $B_1$ and $B_2$ are spanned by orthogonal linear combinations of those, leading to the familiar $2\sqrt{2}$ Tsirelson's bound (for detailed calculation, see Appendix \ref{app:maxcorr}).

While this discussion allows one to explore the role of indistinguishability within quantum mechanics, we may look for more generality beyond quantum theory. Based on sharp measurement formalism for GPTs,  we shall explore the extension of indistinguishability to this broader scenario in the rest of this paper.

{\it Sharp measurements --- } 
For GPTs, following Chiribella  and Yuan \cite{Chiribella:2014cy}, we apply the concept of sharp measurements together with E principle to discuss $n$-body nonlocality. 
Sharp measurements are idealized measurements in GPT formalisms that are minimally disturbing and repeatable. In other words, a sharp measurement is assumed to be realizable on a system many times (repeatable) without influencing the result of any other compatible measurement (minimally disturbing). Also, it is possible to realize a sharp measurement by joining a measurement (not necessarily sharp) on the system and the environment together. In addition to such definition properties, two other properties arise from simple principles: (a) two sharp measurements taken together also makes a sharp measurement and (b) coarse-graining (i.e., combining various information of) a sharp measurement also gives a sharp measurement (less information, more sharpness principle). These assumptions lead to E principle  (described below) as well as to the exclusive hierarchy \cite{Chiribella:2014cy}.

E principle states that if $n$ events are pairwise exclusive, they must also be $n$-wise exclusive. Two events are said to be exclusive if they cannot occur at the same time, which means that they are disjunct and the summation of their probabilities must be at most unity. It has given an explanation for quantum contextuality\cite{Cabello:2013ej}, and later was also applied to the nonlocality problem\cite{Cabello:2015ka}, as we will discuss in more details later. 

\begin{figure}[tb]
\includegraphics[width=0.45\textwidth]{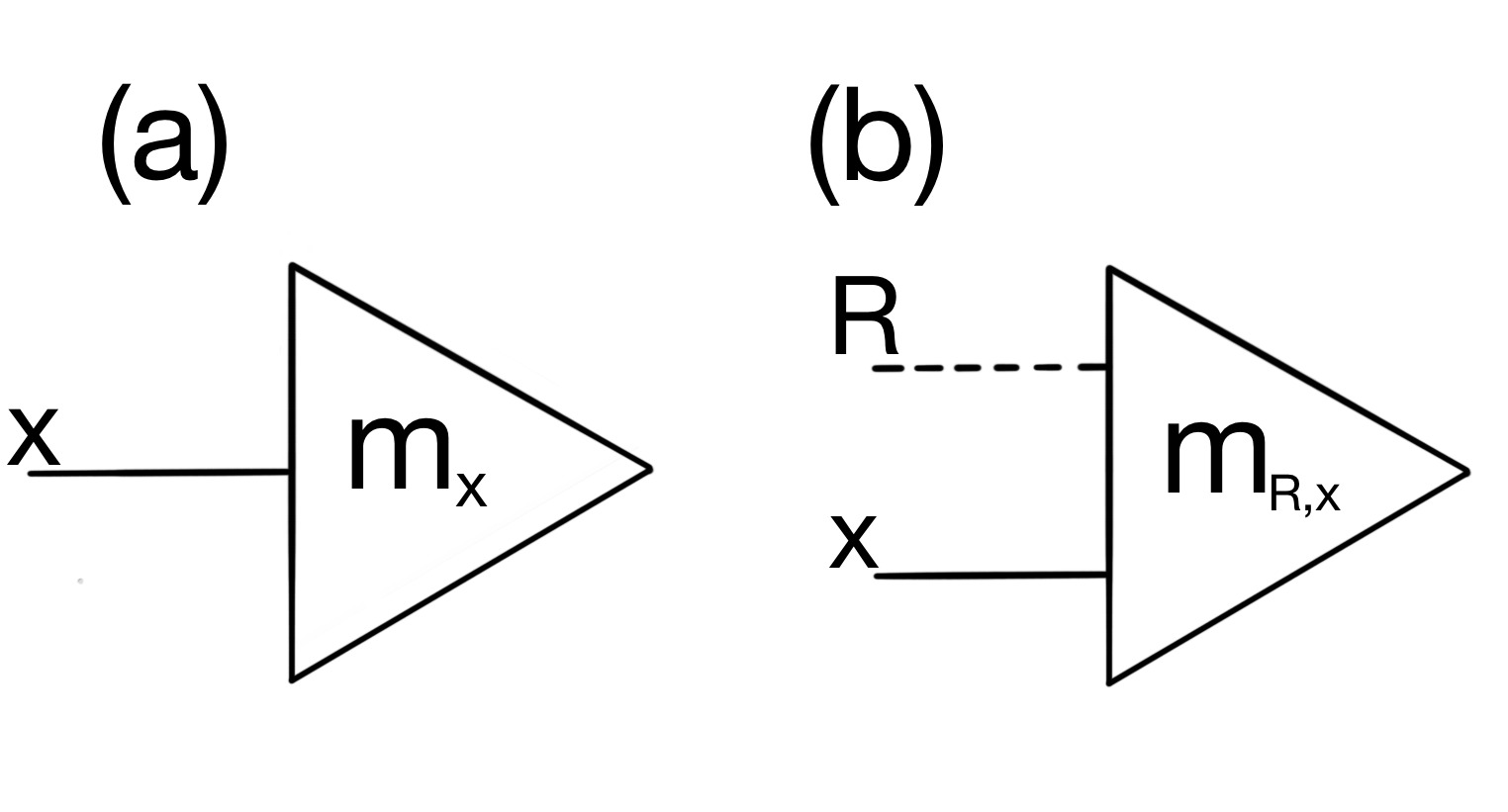}
\caption{(a) Schematic representation of sharp measurement according to ref. \onlinecite{Chiribella:2014cy}. (b) Adjustment of the scheme in (a) by explicitly adding a reference to gauge the measurement, represented by the dashed line.}
\label{sharp}
\end{figure}

To build a GPT footing similar to the mentioned above, we must translate sharp measurement formalism to our concept of physical reference/encoding double bit entry. A general bit (gbit) following such GPTs with sharp measurements shall be represented as $|r,b)$, again having $r$ to stand for its reference bit and $b$ for the coding bit. Accordingly, $(m_{R,x}|$ shall represent effects (i.e. transformations) on such states, with $R$ being the reference input and $x$ denoting relevant bases (fig. \ref{sharp}). The coding information in $b$ is defined regarding reference $r$ as is the effect gauged by some $R$. It differs from ref. \onlinecite{Chiribella:2014cy} by explicitly adding the reference input that is tacitly assumed in the concept of sharp measurements. For a sharp measurement on multiple parties, either one or multiple references may be  present in principle, though not more than one per retrieved information bit, for consistency.

Above mentioned assumptions on joint sharp measurements (a) and coarse-graining (b) can be readily generalized for our approach. Joining measurements $(m_{R,x}|$ and $(n_{R',y}|$ may be represented by writing  $(m_{R,x}|\otimes(n_{R',y}| = (m'_{RR',xy}|$, though the precise method for computing it is irrelevant. Coarse graining may be thought of in two manners: (i)  coding bits with the same physical reference may be coarse-grained  or (ii)  a pair of coding bits and their references may be coarse-grained together in any situation. We will use it to revisit E principle application to $n$-body non-locality.

{\it Information indistinguishability and E principle --- }  
Exclusivity principle states that $n$ pairwise exclusive events (i.e., events that cannot happen simultaneously) are also $n$-wise exclusive, and it places probabilistic constraints to such events (which are disjoint) since their overall summed probability cannot exceed 1. Cabello has used it to show that quantum bounds of $n$-body nonlocality derived by Collins et al. and Seevinck and Svetlichny \cite{Collins:2002ky,*Seevinck:2002hm} can be explained by such probabilistic principle. The inequality in question,
\begin{eqnarray}
S_n \le_{\mathrm{H}} 2^{n-1} \le_{\mathrm{QT}} 2^{n-1}\sqrt{2},
\label{n:body}
\end{eqnarray}
limits classical correlation bounds given by hidden variables (local realism), represented by the subscript ``H'', and the correlation bound for quantum theory, identified by the subscript ``QT.'' $S_n$ can be calculated recursively from the relation $S_n = \langle s_{n-1}x_1\rangle + \langle \bar{s}_{n-1}x_0\rangle $, where $S_2= \langle s_2\rangle = \langle x_0x_0\rangle + \langle x_0x_1\rangle + \langle x_1x_0\rangle - \langle x_1x_1\rangle$ is the CHSH correlation and $\bar{s}$ is obtained by inverting 0s and 1s. 

For the application of E principle, four assumptions must be made. The first three are the assumptions that define sharp measurements, viz. the existence of a sharp measurement on the system and the environment together, joining of sharp measurements into a sharp measurement, and coarse-graining of sharp measurements, for they imply E principle by themselves. A fourth assumption is the existence of a sharp measurement $A_{ij}$ s.t. $A_{00}$ and $A_{11}$ are compatible and $A_{01}$ and $A_{10}$ are also compatible. $A_{ij}$ is composed by sharp measurements $x=0$ and $x=1$ on system $A$ and $x'=0$ and $x'=1$ on system $A'$, yielding 0 if measurements $x=i$ and $x'=j$ return the same result and 1 otherwise. This assumption is used to derive quantum bounds to eq. (\ref{n:body}) from two copies of a system, $A$ and $A'$. This assumption can be justified by information indistinguishability.

Let a system $A$ and a copy $A'$ of it be prepared and represented by $|b_1, b_2, \ldots)$ and $|b'_1, b'_2, \ldots)$ respectively, where the prime becomes the reference bit for the system. Systems $A$ and $A'$ may be located far apart, which is the information indicated by the prime, and are therefore distinguishable in principle. The composite system $AA'$ can then be represented by $|b_1,b'_1; b_2,b'_2;\ldots)$. By definition, $A_{ij}$ is a measurement that does not differentiate gbits $b_1$ and $b'_1$. We may therefore drop the background information signalized by the prime bit, rewriting the state as $|b_1,\tilde{b}_1; b_2,\tilde{b}_2;\ldots)$, where bits $b_i$ and $\tilde{b}_i$ become mutual reference and coding information. This implies equivalence between states $|00)$ and $|11)$ and states $|01)$ and $|10)$. This state equivalence assures that there are at least two compatible measurements to each indistinguishable scenario, namely $A_{00}/A_{11}$ and $A_{01}/A_{10}$. 
This is enough to assure tight quantum bounds in eq. (\ref{n:body}), with a brief derivation given in the Appendix \ref{app:nonlocal}. For a detailed derivation, see ref. \onlinecite{Cabello:2015ka}.

{\it Conclusion and discussion ---} 
Information indistinguishability provides an interpretation that extends the indistinguishability of (identical) particles to information units like qubits or an equivalent binary state in GPTs we refer to as gbits. In quantum theory, it can be understood as a dissociation of encoding internal degrees of freedom of a qubit, rendered indistinguishable if taken alone, and their real physical properties like charge, mass, position and so on that allows qubits to be distinct. By explicitly separating such information into a physical background reference and an internal coding bit, we may ignore such reference when their internal degrees of freedom are coupled and assume their new reference to become one another in a pair of qubits.

In a more general perspective, we can analyze general bits on GPTs by using sharp measurements and E principle implied by it. We observe that information indistinguishability sustains the existence of non-local sharp measurements that may be used to complete inequalities to derive tight quantum bounds for nonlocality. This supports the consideration of information indistinguishability as a fundamental physical principle generator and restrictor of entanglement. Indeed, a Popescu-Rohrlich-like correlation box that extrapolates quantum nonlocality violates information indistinguishability. For example, consider two gbits and two measurements each, where three measurement pairings give perfect correlation and one pairing no correlation at all, hence $S_2=3$. By pinning one axes combination to lack correlation, the particles are identified along these axes, for if we assume that gbits become correlated when they are indistinguishable, the non-correlated axis will allow to track down the correlation process, identifying each party during coupling, contradicting our hypothesis. Such lack of transitivity between perfect correlations (viz. measurements $A_1$ and $A_2$ perfectly correlate with $B_1$,  $B_2$ perfectly correlate with $A_1$, but such correlation does not extend to $A_2$ and $B_2$) is unrealistic and  incompatible with information indistinguishability. This incompatibility suggests the consideration of information indistinguishability as an indicator for realistic theories.

Our approach suggests information indistinguishability as means for nonlocal realistic entanglement, offering a relatively simple understanding for the limitation of realistic nonlocality in quantum theory.

\begin{acknowledgements}
\vspace{1 cm}
The author is thankful to Yuki Kawaguchi for most valuable discussions. This work was supported by JSPS KAKENHI Grants Number JP15J10568.
\end{acknowledgements}

%\nolinenumbers
\bibliography{refs}

\begin{thebibliography}{28}%
\makeatletter
\providecommand \@ifxundefined [1]{%
 \@ifx{#1\undefined}
}%
\providecommand \@ifnum [1]{%
 \ifnum #1\expandafter \@firstoftwo
 \else \expandafter \@secondoftwo
 \fi
}%
\providecommand \@ifx [1]{%
 \ifx #1\expandafter \@firstoftwo
 \else \expandafter \@secondoftwo
 \fi
}%
\providecommand \natexlab [1]{#1}%
\providecommand \enquote  [1]{``#1''}%
\providecommand \bibnamefont  [1]{#1}%
\providecommand \bibfnamefont [1]{#1}%
\providecommand \citenamefont [1]{#1}%
\providecommand \href@noop [0]{\@secondoftwo}%
\providecommand \href [0]{\begingroup \@sanitize@url \@href}%
\providecommand \@href[1]{\@@startlink{#1}\@@href}%
\providecommand \@@href[1]{\endgroup#1\@@endlink}%
\providecommand \@sanitize@url [0]{\catcode `\\12\catcode `\$12\catcode
  `\&12\catcode `\#12\catcode `\^12\catcode `\_12\catcode `\%12\relax}%
\providecommand \@@startlink[1]{}%
\providecommand \@@endlink[0]{}%
\providecommand \url  [0]{\begingroup\@sanitize@url \@url }%
\providecommand \@url [1]{\endgroup\@href {#1}{\urlprefix }}%
\providecommand \urlprefix  [0]{URL }%
\providecommand \Eprint [0]{\href }%
\providecommand \doibase [0]{http://dx.doi.org/}%
\providecommand \selectlanguage [0]{\@gobble}%
\providecommand \bibinfo  [0]{\@secondoftwo}%
\providecommand \bibfield  [0]{\@secondoftwo}%
\providecommand \translation [1]{[#1]}%
\providecommand \BibitemOpen [0]{}%
\providecommand \bibitemStop [0]{}%
\providecommand \bibitemNoStop [0]{.\EOS\space}%
\providecommand \EOS [0]{\spacefactor3000\relax}%
\providecommand \BibitemShut  [1]{\csname bibitem#1\endcsname}%
\let\auto@bib@innerbib\@empty
%</preamble>
\bibitem [{\citenamefont {Einstein}\ \emph {et~al.}(1935)\citenamefont
  {Einstein}, \citenamefont {Podolsky},\ and\ \citenamefont
  {Rosen}}]{Einstein:1935rr}%
  \BibitemOpen
  \bibfield  {author} {\bibinfo {author} {\bibfnamefont {Albert}\ \bibnamefont
  {Einstein}}, \bibinfo {author} {\bibfnamefont {Boris}\ \bibnamefont
  {Podolsky}}, \ and\ \bibinfo {author} {\bibfnamefont {Nathan}\ \bibnamefont
  {Rosen}},\ }\bibfield  {title} {\enquote {\bibinfo {title} {{Can quantum
  mechanical description of physical reality be considered complete?}}}\
  }\href@noop {} {\bibfield  {journal} {\bibinfo  {journal} {Phys. Rev.}\
  }\textbf {\bibinfo {volume} {47}},\ \bibinfo {pages} {777--780} (\bibinfo
  {year} {1935})}\BibitemShut {NoStop}%
\bibitem [{Note1()}]{Note1}%
  \BibitemOpen
  \bibinfo {note} {By ``realism'' we assume the possibility of measuring some
  physical property, local or non-local, unambiguously, that is also
  confirmable by other measurements. A Popescu-Rohrllich box, which breaks
  translation of perfect correlations and is discussed later, cannot be said
  realistic in these terms, for while a set of measurements can be combined to
  lead to the same parity result, one combination gives an opposite,
  contradictory result for this same physical property.}\BibitemShut {Stop}%
\bibitem [{\citenamefont {Bell}(1964)}]{Bell:1964wu}%
  \BibitemOpen
  \bibfield  {author} {\bibinfo {author} {\bibfnamefont {J~S}\ \bibnamefont
  {Bell}},\ }\bibfield  {title} {\enquote {\bibinfo {title} {{On the Einstein
  Podolsky Rosen paradox}},}\ }\href@noop {} {\bibfield  {journal} {\bibinfo
  {journal} {Physics}\ }\textbf {\bibinfo {volume} {1}},\ \bibinfo {pages}
  {195--290} (\bibinfo {year} {1964})}\BibitemShut {NoStop}%
\bibitem [{\citenamefont {Clauser}\ \emph {et~al.}(1969)\citenamefont
  {Clauser}, \citenamefont {Horne}, \citenamefont {Shimony},\ and\
  \citenamefont {Holt}}]{Clauser:1969ny}%
  \BibitemOpen
  \bibfield  {author} {\bibinfo {author} {\bibfnamefont {John~F}\ \bibnamefont
  {Clauser}}, \bibinfo {author} {\bibfnamefont {Michael~A}\ \bibnamefont
  {Horne}}, \bibinfo {author} {\bibfnamefont {Abner}\ \bibnamefont {Shimony}},
  \ and\ \bibinfo {author} {\bibfnamefont {Richard~A}\ \bibnamefont {Holt}},\
  }\bibfield  {title} {\enquote {\bibinfo {title} {{Proposed experiment to test
  local hidden variable theories}},}\ }\href@noop {} {\bibfield  {journal}
  {\bibinfo  {journal} {Physical Review Letters}\ }\textbf {\bibinfo {volume}
  {23}},\ \bibinfo {pages} {880--884} (\bibinfo {year} {1969})}\BibitemShut
  {NoStop}%
\bibitem [{\citenamefont {Mermin}(1993)}]{Mermin:1993di}%
  \BibitemOpen
  \bibfield  {author} {\bibinfo {author} {\bibfnamefont {N~David}\ \bibnamefont
  {Mermin}},\ }\bibfield  {title} {\enquote {\bibinfo {title} {{Hidden
  variables and the two theorems of John Bell}},}\ }\href@noop {} {\bibfield
  {journal} {\bibinfo  {journal} {Reviews of Modern Physics}\ }\textbf
  {\bibinfo {volume} {65}},\ \bibinfo {pages} {803--815} (\bibinfo {year}
  {1993})}\BibitemShut {NoStop}%
\bibitem [{\citenamefont {Peres}(1993)}]{Peres:1993ti}%
  \BibitemOpen
  \bibfield  {author} {\bibinfo {author} {\bibfnamefont {Asher}\ \bibnamefont
  {Peres}},\ }\href@noop {} {\emph {\bibinfo {title} {{Quantum Theory: Concepts
  and Methods}}}},\ edited by\ \bibinfo {editor} {\bibfnamefont {Alwyn}\
  \bibnamefont {van~der Merwe}},\ Vol.~\bibinfo {volume} {57}\ (\bibinfo
  {publisher} {Kluwer Academic Publishers},\ \bibinfo {year}
  {1993})\BibitemShut {NoStop}%
\bibitem [{\citenamefont {Horodecki}\ \emph {et~al.}(2009)\citenamefont
  {Horodecki}, \citenamefont {Horodecki}, \citenamefont {Horodecki},\ and\
  \citenamefont {Horodecki}}]{Horodecki:2009gb}%
  \BibitemOpen
  \bibfield  {author} {\bibinfo {author} {\bibfnamefont {Ryszard}\ \bibnamefont
  {Horodecki}}, \bibinfo {author} {\bibfnamefont {Pawe{\l}}\ \bibnamefont
  {Horodecki}}, \bibinfo {author} {\bibfnamefont {Micha{\l}}\ \bibnamefont
  {Horodecki}}, \ and\ \bibinfo {author} {\bibfnamefont {Karol}\ \bibnamefont
  {Horodecki}},\ }\bibfield  {title} {\enquote {\bibinfo {title} {{Quantum
  entanglement}},}\ }\href@noop {} {\bibfield  {journal} {\bibinfo  {journal}
  {Reviews of Modern Physics}\ }\textbf {\bibinfo {volume} {81}},\ \bibinfo
  {pages} {865--942} (\bibinfo {year} {2009})}\BibitemShut {NoStop}%
\bibitem [{\citenamefont {Giustina}\ \emph {et~al.}(2015)\citenamefont
  {Giustina}, \citenamefont {Versteegh}, \citenamefont {Wengerowsky},
  \citenamefont {Handsteiner}, \citenamefont {Hochrainer}, \citenamefont
  {Phelan}, \citenamefont {Steinlechner}, \citenamefont {Kofler}, \citenamefont
  {Larsson}, \citenamefont {Abell{\'a}n}, \citenamefont {Amaya}, \citenamefont
  {Pruneri}, \citenamefont {Mitchell}, \citenamefont {Beyer}, \citenamefont
  {Gerrits}, \citenamefont {Lita}, \citenamefont {Shalm}, \citenamefont {Nam},
  \citenamefont {Scheidl}, \citenamefont {Ursin}, \citenamefont {Wittmann},\
  and\ \citenamefont {Zeilinger}}]{Giustina:2015fc}%
  \BibitemOpen
  \bibfield  {author} {\bibinfo {author} {\bibfnamefont {Marissa}\ \bibnamefont
  {Giustina}}, \bibinfo {author} {\bibfnamefont {Marijn A~M}\ \bibnamefont
  {Versteegh}}, \bibinfo {author} {\bibfnamefont {S{\"o}ren}\ \bibnamefont
  {Wengerowsky}}, \bibinfo {author} {\bibfnamefont {Johannes}\ \bibnamefont
  {Handsteiner}}, \bibinfo {author} {\bibfnamefont {Armin}\ \bibnamefont
  {Hochrainer}}, \bibinfo {author} {\bibfnamefont {Kevin}\ \bibnamefont
  {Phelan}}, \bibinfo {author} {\bibfnamefont {Fabian}\ \bibnamefont
  {Steinlechner}}, \bibinfo {author} {\bibfnamefont {Johannes}\ \bibnamefont
  {Kofler}}, \bibinfo {author} {\bibfnamefont {Jan-{\AA}ke}\ \bibnamefont
  {Larsson}}, \bibinfo {author} {\bibfnamefont {Carlos}\ \bibnamefont
  {Abell{\'a}n}}, \bibinfo {author} {\bibfnamefont {Waldimar}\ \bibnamefont
  {Amaya}}, \bibinfo {author} {\bibfnamefont {Valerio}\ \bibnamefont
  {Pruneri}}, \bibinfo {author} {\bibfnamefont {Morgan~W}\ \bibnamefont
  {Mitchell}}, \bibinfo {author} {\bibfnamefont {J{\"o}rn}\ \bibnamefont
  {Beyer}}, \bibinfo {author} {\bibfnamefont {Thomas}\ \bibnamefont {Gerrits}},
  \bibinfo {author} {\bibfnamefont {Adriana~E}\ \bibnamefont {Lita}}, \bibinfo
  {author} {\bibfnamefont {Lynden~K}\ \bibnamefont {Shalm}}, \bibinfo {author}
  {\bibfnamefont {Sae~Woo}\ \bibnamefont {Nam}}, \bibinfo {author}
  {\bibfnamefont {Thomas}\ \bibnamefont {Scheidl}}, \bibinfo {author}
  {\bibfnamefont {Rupert}\ \bibnamefont {Ursin}}, \bibinfo {author}
  {\bibfnamefont {Bernhard}\ \bibnamefont {Wittmann}}, \ and\ \bibinfo {author}
  {\bibfnamefont {Anton}\ \bibnamefont {Zeilinger}},\ }\bibfield  {title}
  {\enquote {\bibinfo {title} {{Significant-Loophole-Free Test of
  Bell{\textquoteright}s Theorem with Entangled Photons}},}\ }\href@noop {}
  {\bibfield  {journal} {\bibinfo  {journal} {Physical Review Letters}\
  }\textbf {\bibinfo {volume} {115}},\ \bibinfo {pages} {250401} (\bibinfo
  {year} {2015})}\BibitemShut {NoStop}%
\bibitem [{\citenamefont {Shalm}\ \emph {et~al.}(2015)\citenamefont {Shalm},
  \citenamefont {Meyer-Scott}, \citenamefont {Christensen}, \citenamefont
  {Bierhorst}, \citenamefont {Wayne}, \citenamefont {Stevens}, \citenamefont
  {Gerrits}, \citenamefont {Glancy}, \citenamefont {Hamel}, \citenamefont
  {Allman}, \citenamefont {Coakley}, \citenamefont {Dyer}, \citenamefont
  {Hodge}, \citenamefont {Lita}, \citenamefont {Verma}, \citenamefont
  {Lambrocco}, \citenamefont {Tortorici}, \citenamefont {Migdall},
  \citenamefont {Zhang}, \citenamefont {Kumor}, \citenamefont {Farr},
  \citenamefont {Marsili}, \citenamefont {Shaw}, \citenamefont {Stern},
  \citenamefont {Abell{\'a}n}, \citenamefont {Amaya}, \citenamefont {Pruneri},
  \citenamefont {Jennewein}, \citenamefont {Mitchell}, \citenamefont {Kwiat},
  \citenamefont {Bienfang}, \citenamefont {Mirin}, \citenamefont {Knill},\ and\
  \citenamefont {Nam}}]{Shalm:2015cw}%
  \BibitemOpen
  \bibfield  {author} {\bibinfo {author} {\bibfnamefont {Lynden~K}\
  \bibnamefont {Shalm}}, \bibinfo {author} {\bibfnamefont {Evan}\ \bibnamefont
  {Meyer-Scott}}, \bibinfo {author} {\bibfnamefont {Bradley~G}\ \bibnamefont
  {Christensen}}, \bibinfo {author} {\bibfnamefont {Peter}\ \bibnamefont
  {Bierhorst}}, \bibinfo {author} {\bibfnamefont {Michael~A}\ \bibnamefont
  {Wayne}}, \bibinfo {author} {\bibfnamefont {Martin~J}\ \bibnamefont
  {Stevens}}, \bibinfo {author} {\bibfnamefont {Thomas}\ \bibnamefont
  {Gerrits}}, \bibinfo {author} {\bibfnamefont {Scott}\ \bibnamefont {Glancy}},
  \bibinfo {author} {\bibfnamefont {Deny~R}\ \bibnamefont {Hamel}}, \bibinfo
  {author} {\bibfnamefont {Michael~S}\ \bibnamefont {Allman}}, \bibinfo
  {author} {\bibfnamefont {Kevin~J}\ \bibnamefont {Coakley}}, \bibinfo {author}
  {\bibfnamefont {Shellee~D}\ \bibnamefont {Dyer}}, \bibinfo {author}
  {\bibfnamefont {Carson}\ \bibnamefont {Hodge}}, \bibinfo {author}
  {\bibfnamefont {Adriana~E}\ \bibnamefont {Lita}}, \bibinfo {author}
  {\bibfnamefont {Varun~B}\ \bibnamefont {Verma}}, \bibinfo {author}
  {\bibfnamefont {Camilla}\ \bibnamefont {Lambrocco}}, \bibinfo {author}
  {\bibfnamefont {Edward}\ \bibnamefont {Tortorici}}, \bibinfo {author}
  {\bibfnamefont {Alan~L}\ \bibnamefont {Migdall}}, \bibinfo {author}
  {\bibfnamefont {Yanbao}\ \bibnamefont {Zhang}}, \bibinfo {author}
  {\bibfnamefont {Daniel~R}\ \bibnamefont {Kumor}}, \bibinfo {author}
  {\bibfnamefont {William~H}\ \bibnamefont {Farr}}, \bibinfo {author}
  {\bibfnamefont {Francesco}\ \bibnamefont {Marsili}}, \bibinfo {author}
  {\bibfnamefont {Matthew~D}\ \bibnamefont {Shaw}}, \bibinfo {author}
  {\bibfnamefont {Jeffrey~A}\ \bibnamefont {Stern}}, \bibinfo {author}
  {\bibfnamefont {Carlos}\ \bibnamefont {Abell{\'a}n}}, \bibinfo {author}
  {\bibfnamefont {Waldimar}\ \bibnamefont {Amaya}}, \bibinfo {author}
  {\bibfnamefont {Valerio}\ \bibnamefont {Pruneri}}, \bibinfo {author}
  {\bibfnamefont {Thomas}\ \bibnamefont {Jennewein}}, \bibinfo {author}
  {\bibfnamefont {Morgan~W}\ \bibnamefont {Mitchell}}, \bibinfo {author}
  {\bibfnamefont {Paul~G}\ \bibnamefont {Kwiat}}, \bibinfo {author}
  {\bibfnamefont {Joshua~C}\ \bibnamefont {Bienfang}}, \bibinfo {author}
  {\bibfnamefont {Richard~P}\ \bibnamefont {Mirin}}, \bibinfo {author}
  {\bibfnamefont {Emanuel}\ \bibnamefont {Knill}}, \ and\ \bibinfo {author}
  {\bibfnamefont {Sae~Woo}\ \bibnamefont {Nam}},\ }\bibfield  {title} {\enquote
  {\bibinfo {title} {{Strong Loophole-Free Test of Local Realism}},}\
  }\href@noop {} {\bibfield  {journal} {\bibinfo  {journal} {Physical Review
  Letters}\ }\textbf {\bibinfo {volume} {115}},\ \bibinfo {pages} {250402}
  (\bibinfo {year} {2015})}\BibitemShut {NoStop}%
\bibitem [{\citenamefont {Hensen}\ \emph {et~al.}(2015)\citenamefont {Hensen},
  \citenamefont {Bernien}, \citenamefont {Dr{\'e}au}, \citenamefont {Reiserer},
  \citenamefont {Kalb}, \citenamefont {Blok}, \citenamefont {Ruitenberg},
  \citenamefont {Vermeulen}, \citenamefont {Schouten}, \citenamefont
  {Abell{\'a}n}, \citenamefont {Amaya}, \citenamefont {Pruneri}, \citenamefont
  {Mitchell}, \citenamefont {Markham}, \citenamefont {Twitchen}, \citenamefont
  {Elkouss}, \citenamefont {Wehner}, \citenamefont {Taminiau},\ and\
  \citenamefont {Hanson}}]{Hensen:2015dw}%
  \BibitemOpen
  \bibfield  {author} {\bibinfo {author} {\bibfnamefont {B}~\bibnamefont
  {Hensen}}, \bibinfo {author} {\bibfnamefont {H}~\bibnamefont {Bernien}},
  \bibinfo {author} {\bibfnamefont {A~E}\ \bibnamefont {Dr{\'e}au}}, \bibinfo
  {author} {\bibfnamefont {A}~\bibnamefont {Reiserer}}, \bibinfo {author}
  {\bibfnamefont {N}~\bibnamefont {Kalb}}, \bibinfo {author} {\bibfnamefont
  {M~S}\ \bibnamefont {Blok}}, \bibinfo {author} {\bibfnamefont
  {J}~\bibnamefont {Ruitenberg}}, \bibinfo {author} {\bibfnamefont {R~F~L}\
  \bibnamefont {Vermeulen}}, \bibinfo {author} {\bibfnamefont {R~N}\
  \bibnamefont {Schouten}}, \bibinfo {author} {\bibfnamefont {C}~\bibnamefont
  {Abell{\'a}n}}, \bibinfo {author} {\bibfnamefont {W}~\bibnamefont {Amaya}},
  \bibinfo {author} {\bibfnamefont {V}~\bibnamefont {Pruneri}}, \bibinfo
  {author} {\bibfnamefont {M~W}\ \bibnamefont {Mitchell}}, \bibinfo {author}
  {\bibfnamefont {M}~\bibnamefont {Markham}}, \bibinfo {author} {\bibfnamefont
  {D~J}\ \bibnamefont {Twitchen}}, \bibinfo {author} {\bibfnamefont
  {D}~\bibnamefont {Elkouss}}, \bibinfo {author} {\bibfnamefont
  {S}~\bibnamefont {Wehner}}, \bibinfo {author} {\bibfnamefont {T~H}\
  \bibnamefont {Taminiau}}, \ and\ \bibinfo {author} {\bibfnamefont
  {R}~\bibnamefont {Hanson}},\ }\bibfield  {title} {\enquote {\bibinfo {title}
  {{Loophole-free Bell inequality violation using electron spins separated by
  1.3 kilometres}},}\ }\href@noop {} {\bibfield  {journal} {\bibinfo  {journal}
  {Nature}\ }\textbf {\bibinfo {volume} {526}},\ \bibinfo {pages} {682--686}
  (\bibinfo {year} {2015})}\BibitemShut {NoStop}%
\bibitem [{\citenamefont {Popescu}\ and\ \citenamefont
  {Rohrlich}(1994)}]{Popescu:1994ku}%
  \BibitemOpen
  \bibfield  {author} {\bibinfo {author} {\bibfnamefont {Sandu}\ \bibnamefont
  {Popescu}}\ and\ \bibinfo {author} {\bibfnamefont {Daniel}\ \bibnamefont
  {Rohrlich}},\ }\bibfield  {title} {\enquote {\bibinfo {title} {{Quantum
  nonlocality as an axiom}},}\ }\href@noop {} {\bibfield  {journal} {\bibinfo
  {journal} {Foundations of Physics}\ }\textbf {\bibinfo {volume} {24}},\
  \bibinfo {pages} {379--385} (\bibinfo {year} {1994})}\BibitemShut {NoStop}%
\bibitem [{\citenamefont {Paw{\l}owski}\ \emph {et~al.}(2009)\citenamefont
  {Paw{\l}owski}, \citenamefont {Paterek}, \citenamefont {Kaszlikowski},
  \citenamefont {Scarani}, \citenamefont {Winter},\ and\ \citenamefont
  {{\.{Z}}ukowski}}]{Pawiowski:2009dt}%
  \BibitemOpen
  \bibfield  {author} {\bibinfo {author} {\bibfnamefont {Marcin}\ \bibnamefont
  {Paw{\l}owski}}, \bibinfo {author} {\bibfnamefont {Tomasz}\ \bibnamefont
  {Paterek}}, \bibinfo {author} {\bibfnamefont {Dagomir}\ \bibnamefont
  {Kaszlikowski}}, \bibinfo {author} {\bibfnamefont {Valerio}\ \bibnamefont
  {Scarani}}, \bibinfo {author} {\bibfnamefont {Andreas}\ \bibnamefont
  {Winter}}, \ and\ \bibinfo {author} {\bibfnamefont {Marek}\ \bibnamefont
  {{\.{Z}}ukowski}},\ }\bibfield  {title} {\enquote {\bibinfo {title}
  {{Information causality as a physical principle}},}\ }\href@noop {}
  {\bibfield  {journal} {\bibinfo  {journal} {Nature}\ }\textbf {\bibinfo
  {volume} {461}},\ \bibinfo {pages} {1101--1104} (\bibinfo {year}
  {2009})}\BibitemShut {NoStop}%
\bibitem [{\citenamefont {Navascues}\ and\ \citenamefont
  {Wunderlich}(2010)}]{Navascues:2010fk}%
  \BibitemOpen
  \bibfield  {author} {\bibinfo {author} {\bibfnamefont {M}~\bibnamefont
  {Navascues}}\ and\ \bibinfo {author} {\bibfnamefont {H}~\bibnamefont
  {Wunderlich}},\ }\bibfield  {title} {\enquote {\bibinfo {title} {{A glance
  beyond the quantum model}},}\ }\href@noop {} {\bibfield  {journal} {\bibinfo
  {journal} {Proceedings of the Royal Society A: Mathematical, Physical and
  Engineering Sciences}\ }\textbf {\bibinfo {volume} {466}},\ \bibinfo {pages}
  {881--890} (\bibinfo {year} {2010})}\BibitemShut {NoStop}%
\bibitem [{\citenamefont {Cabello}(2015)}]{Cabello:2015ka}%
  \BibitemOpen
  \bibfield  {author} {\bibinfo {author} {\bibfnamefont {Ad{\'a}n}\
  \bibnamefont {Cabello}},\ }\bibfield  {title} {\enquote {\bibinfo {title}
  {{Simple Explanation of the Quantum Limits of Genuine n-Body Nonlocality}},}\
  }\href@noop {} {\bibfield  {journal} {\bibinfo  {journal} {Physical Review
  Letters}\ }\textbf {\bibinfo {volume} {114}},\ \bibinfo {pages} {220402}
  (\bibinfo {year} {2015})}\BibitemShut {NoStop}%
\bibitem [{\citenamefont {Collins}\ \emph {et~al.}(2002)\citenamefont
  {Collins}, \citenamefont {Gisin}, \citenamefont {Popescu}, \citenamefont
  {Roberts},\ and\ \citenamefont {Scarani}}]{Collins:2002ky}%
  \BibitemOpen
  \bibfield  {author} {\bibinfo {author} {\bibfnamefont {Daniel}\ \bibnamefont
  {Collins}}, \bibinfo {author} {\bibfnamefont {Nicolas}\ \bibnamefont
  {Gisin}}, \bibinfo {author} {\bibfnamefont {Sandu}\ \bibnamefont {Popescu}},
  \bibinfo {author} {\bibfnamefont {David}\ \bibnamefont {Roberts}}, \ and\
  \bibinfo {author} {\bibfnamefont {Valerio}\ \bibnamefont {Scarani}},\
  }\bibfield  {title} {\enquote {\bibinfo {title} {{Bell-Type Inequalities to
  Detect True n-Body Nonseparability}},}\ }\href@noop {} {\bibfield  {journal}
  {\bibinfo  {journal} {Physical Review Letters}\ }\textbf {\bibinfo {volume}
  {88}},\ \bibinfo {pages} {170405} (\bibinfo {year} {2002})}\BibitemShut
  {NoStop}%
\bibitem [{\citenamefont {Seevinck}\ and\ \citenamefont
  {Svetlichny}(2002)}]{Seevinck:2002hm}%
  \BibitemOpen
  \bibfield  {author} {\bibinfo {author} {\bibfnamefont {Michael}\ \bibnamefont
  {Seevinck}}\ and\ \bibinfo {author} {\bibfnamefont {George}\ \bibnamefont
  {Svetlichny}},\ }\bibfield  {title} {\enquote {\bibinfo {title} {{Bell-Type
  Inequalities for Partial Separability in N-Particle Systems and Quantum
  Mechanical Violations}},}\ }\href@noop {} {\bibfield  {journal} {\bibinfo
  {journal} {Physical Review Letters}\ }\textbf {\bibinfo {volume} {89}},\
  \bibinfo {pages} {032112} (\bibinfo {year} {2002})}\BibitemShut {NoStop}%
\bibitem [{\citenamefont {Lo~Franco}\ and\ \citenamefont
  {Compagno}(2016)}]{LoFranco:2016jk}%
  \BibitemOpen
  \bibfield  {author} {\bibinfo {author} {\bibfnamefont {Rosario}\ \bibnamefont
  {Lo~Franco}}\ and\ \bibinfo {author} {\bibfnamefont {Giuseppe}\ \bibnamefont
  {Compagno}},\ }\bibfield  {title} {\enquote {\bibinfo {title} {{Quantum
  entanglement of identical particles by standard information-theoretic
  notions}},}\ }\href@noop {} {\bibfield  {journal} {\bibinfo  {journal}
  {Scientific Reports}\ }\textbf {\bibinfo {volume} {6}},\ \bibinfo {pages}
  {20603} (\bibinfo {year} {2016})}\BibitemShut {NoStop}%
\bibitem [{\citenamefont {Sciara}\ \emph {et~al.}(2017)\citenamefont {Sciara},
  \citenamefont {Lo~Franco},\ and\ \citenamefont {Compagno}}]{Sciara:2017cz}%
  \BibitemOpen
  \bibfield  {author} {\bibinfo {author} {\bibfnamefont {Stefania}\
  \bibnamefont {Sciara}}, \bibinfo {author} {\bibfnamefont {Rosario}\
  \bibnamefont {Lo~Franco}}, \ and\ \bibinfo {author} {\bibfnamefont
  {Giuseppe}\ \bibnamefont {Compagno}},\ }\bibfield  {title} {\enquote
  {\bibinfo {title} {{Universality of Schmidt decomposition and particle
  identity}},}\ }\href@noop {} {\bibfield  {journal} {\bibinfo  {journal}
  {Scientific Reports}\ }\textbf {\bibinfo {volume} {7}},\ \bibinfo {pages}
  {44675} (\bibinfo {year} {2017})}\BibitemShut {NoStop}%
\bibitem [{\citenamefont {Sasaki}\ \emph {et~al.}(2011)\citenamefont {Sasaki},
  \citenamefont {Ichikawa},\ and\ \citenamefont {Tsutsui}}]{Sasaki:2011hk}%
  \BibitemOpen
  \bibfield  {author} {\bibinfo {author} {\bibfnamefont {Toshihiko}\
  \bibnamefont {Sasaki}}, \bibinfo {author} {\bibfnamefont {Tsubasa}\
  \bibnamefont {Ichikawa}}, \ and\ \bibinfo {author} {\bibfnamefont {Izumi}\
  \bibnamefont {Tsutsui}},\ }\bibfield  {title} {\enquote {\bibinfo {title}
  {{Entanglement of indistinguishable particles}},}\ }\href@noop {} {\bibfield
  {journal} {\bibinfo  {journal} {Physical Review A}\ }\textbf {\bibinfo
  {volume} {83}},\ \bibinfo {pages} {012113} (\bibinfo {year}
  {2011})}\BibitemShut {NoStop}%
\bibitem [{\citenamefont {Eckert}\ \emph {et~al.}(2002)\citenamefont {Eckert},
  \citenamefont {Schliemann}, \citenamefont {Bruss},\ and\ \citenamefont
  {Lewenstein}}]{Eckert:2002ij}%
  \BibitemOpen
  \bibfield  {author} {\bibinfo {author} {\bibfnamefont {K}~\bibnamefont
  {Eckert}}, \bibinfo {author} {\bibfnamefont {J}~\bibnamefont {Schliemann}},
  \bibinfo {author} {\bibfnamefont {D}~\bibnamefont {Bruss}}, \ and\ \bibinfo
  {author} {\bibfnamefont {M}~\bibnamefont {Lewenstein}},\ }\bibfield  {title}
  {\enquote {\bibinfo {title} {{Quantum Correlations in Systems of
  Indistinguishable Particles}},}\ }\href@noop {} {\bibfield  {journal}
  {\bibinfo  {journal} {Annals of Physics}\ }\textbf {\bibinfo {volume}
  {299}},\ \bibinfo {pages} {88--127} (\bibinfo {year} {2002})}\BibitemShut
  {NoStop}%
\bibitem [{\citenamefont {Cirel'son}(1980)}]{Cirelson:1980fp}%
  \BibitemOpen
  \bibfield  {author} {\bibinfo {author} {\bibfnamefont {B~S}\ \bibnamefont
  {Cirel'son}},\ }\bibfield  {title} {\enquote {\bibinfo {title} {{Quantum
  generalizations of Bell's inequality}},}\ }\href@noop {} {\bibfield
  {journal} {\bibinfo  {journal} {Letters in Mathematical Physics}\ }\textbf
  {\bibinfo {volume} {4}},\ \bibinfo {pages} {93--100} (\bibinfo {year}
  {1980})}\BibitemShut {NoStop}%
\bibitem [{\citenamefont {Chiribella}\ and\ \citenamefont
  {Yuan}(2014)}]{Chiribella:2014cy}%
  \BibitemOpen
  \bibfield  {author} {\bibinfo {author} {\bibfnamefont {G}~\bibnamefont
  {Chiribella}}\ and\ \bibinfo {author} {\bibfnamefont {X}~\bibnamefont
  {Yuan}},\ }\bibfield  {title} {\enquote {\bibinfo {title} {{Measurement
  sharpness cuts nonlocality and contextuality in every physical theory}},}\
  }\href@noop {} {\bibfield  {journal} {\bibinfo  {journal} {arXiv.org}\ ,\
  \bibinfo {pages} {1--14}} (\bibinfo {year} {2014})},\ \Eprint
  {http://arxiv.org/abs/1404.3348v2} {1404.3348v2} \BibitemShut {NoStop}%
\bibitem [{\citenamefont {Ivanovic}(1987)}]{Ivanovic:1987bh}%
  \BibitemOpen
  \bibfield  {author} {\bibinfo {author} {\bibfnamefont {I~D}\ \bibnamefont
  {Ivanovic}},\ }\bibfield  {title} {\enquote {\bibinfo {title} {{How to
  differentiate between non-orthogonal states}},}\ }\href@noop {} {\bibfield
  {journal} {\bibinfo  {journal} {Physics Letters A}\ }\textbf {\bibinfo
  {volume} {123}},\ \bibinfo {pages} {257--259} (\bibinfo {year}
  {1987})}\BibitemShut {NoStop}%
\bibitem [{\citenamefont {Dieks}(1988)}]{Dieks:1988iy}%
  \BibitemOpen
  \bibfield  {author} {\bibinfo {author} {\bibfnamefont {D}~\bibnamefont
  {Dieks}},\ }\bibfield  {title} {\enquote {\bibinfo {title} {{Overlap and
  distinguishability of quantum states}},}\ }\href@noop {} {\bibfield
  {journal} {\bibinfo  {journal} {Physics Letters A}\ }\textbf {\bibinfo
  {volume} {126}},\ \bibinfo {pages} {303--306} (\bibinfo {year}
  {1988})}\BibitemShut {NoStop}%
\bibitem [{\citenamefont {Kawakubo}\ and\ \citenamefont
  {Koike}(2016)}]{Kawakubo:2016jd}%
  \BibitemOpen
  \bibfield  {author} {\bibinfo {author} {\bibfnamefont {Ry{\^u}itir{\^o}}\
  \bibnamefont {Kawakubo}}\ and\ \bibinfo {author} {\bibfnamefont {Tatsuhiko}\
  \bibnamefont {Koike}},\ }\bibfield  {title} {\enquote {\bibinfo {title}
  {{Distinguishability of countable quantum states and von Neumann lattice}},}\
  }\href@noop {} {\bibfield  {journal} {\bibinfo  {journal} {Journal of Physics
  A: Mathematical and Theoretical}\ }\textbf {\bibinfo {volume} {49}},\
  \bibinfo {pages} {265201} (\bibinfo {year} {2016})}\BibitemShut {NoStop}%
\bibitem [{\citenamefont {Cabello}(2013)}]{Cabello:2013ej}%
  \BibitemOpen
  \bibfield  {author} {\bibinfo {author} {\bibfnamefont {Ad{\'a}n}\
  \bibnamefont {Cabello}},\ }\bibfield  {title} {\enquote {\bibinfo {title}
  {{Simple Explanation of the Quantum Violation of a Fundamental
  Inequality}},}\ }\href@noop {} {\bibfield  {journal} {\bibinfo  {journal}
  {Physical Review Letters}\ }\textbf {\bibinfo {volume} {110}},\ \bibinfo
  {pages} {060402} (\bibinfo {year} {2013})}\BibitemShut {NoStop}%
\bibitem [{\citenamefont {Dym}(2007)}]{Dym:2007ug}%
  \BibitemOpen
  \bibfield  {author} {\bibinfo {author} {\bibfnamefont {H}~\bibnamefont
  {Dym}},\ }\href@noop {} {\emph {\bibinfo {title} {{Linear algebra in action,
  volume 78 of Graduate Studies in Mathematics}}}}\ (\bibinfo  {publisher}
  {American Mathematical Society},\ \bibinfo {year} {2007})\BibitemShut
  {NoStop}%
\bibitem [{\citenamefont {Nielsen}\ and\ \citenamefont
  {Chuang}(2010)}]{Nielsen:2010vn}%
  \BibitemOpen
  \bibfield  {author} {\bibinfo {author} {\bibfnamefont {Michael~A}\
  \bibnamefont {Nielsen}}\ and\ \bibinfo {author} {\bibfnamefont {Isaac~L}\
  \bibnamefont {Chuang}},\ }\href@noop {} {\emph {\bibinfo {title} {{Quantum
  Computation and Quantum Information}}}},\ 10th Anniversary Edition\ (\bibinfo
   {publisher} {Cambridge University Press},\ \bibinfo {year}
  {2010})\BibitemShut {NoStop}%
\end{thebibliography}%

\appendix
%\section{Appendix}
%\section{Tsirelson's Bound}

%A brief proof of Tsirelson's bound is given following Peres \cite{Peres:1993ti}. We write
%\begin{eqnarray}
%C = \alpha\beta + \gamma\beta + \alpha\delta - \gamma\delta,
%\end{eqnarray}
%assuming they either commute or anti-commute, i.e. $[\alpha,\beta]_{\pm} = [\beta, \gamma]_{\pm} = [\gamma,\delta]_{\pm} = [\delta,\alpha]_{\pm} = 0$. Hence, one can write
%\begin{eqnarray}
%C^2 &=& \mp 4 + [\alpha,\gamma]_{-}[\beta,\delta]_{-},\\
%||C^2|| &\le& 4 + ||[\alpha,\gamma]|| \,||[\beta,\delta]||\nonumber\\
%||C^2|| &\le& 4 + 4||\alpha\gamma|| \,||\beta\delta||\nonumber\\
%\therefore ||C|| &\le& 2\sqrt{2}.
%\end{eqnarray}
%This can be understood by noticing that when squaring $C$, terms like $\alpha\beta\gamma\beta$ cancel out with terms like $-\alpha\delta\gamma\delta$, whether these operators commute or anti-commute, leaving only a $\alpha\beta\gamma\delta$ term.

\section{Lemma and proof}\label{app:lemma}
\begin{lemma*}
Given two projectors $P$ and $Q$ obeying $|| P - Q || < 1$, $\rank P = \rank Q$.
\label{lemma}
\end{lemma*}

This lemma and proof is provided by Dym\cite{Dym:2007ug}. If $P$ and $Q$ are two projectors  and $|| P-Q || < 1$, clearly $I_n - (P-Q)$ is full rank and invertible. Hence:
\begin{eqnarray}
\rank P &=& \rank P(I_n - (P-Q)) = \rank PQ\nonumber\\
&\le& \rank Q.
\end{eqnarray}
The second equality comes from projectors' idempotency ($P^2=P$). Since the same argument can be made exchanging $P$ and $Q$, they must have the same rank. QED.

\section{Schmidt decomposition}\label{app:SD}
Given a bipartite quantum system, it can in general be represented as

\begin{eqnarray}
|\Psi\rangle &=& \sum_{ij} c_{ij}|i\rangle|j\rangle.
\end{eqnarray}

By performing a singular value decomposition (SVD) of the coefficient matrix, one obtains
\begin{eqnarray}
|\Psi\rangle &=& \sum_{ijk} v_{ik}\sqrt{\lambda_k}u_{kj}|i\rangle|j\rangle\nonumber\\
&=& \sum_k\sqrt{\lambda_k}|\lambda_k^A\rangle|\lambda_k^B\rangle,
\end{eqnarray}
where $\sqrt{\lambda_k}$ are the singular values, and the correlated kets in the second line are obtained by applying $v_{ik}$ ($u_{kj}$) to $|i\rangle$ ($|j\rangle$). These are called Schmidt bases, and the rank of the matrix $\Lambda = (\sqrt{\lambda_k})$, i.e., the number of singular values, is called Schmidt rank. A Schmidt rank equal to 1 implies that the quantum state is separable into the product of two independent states, and is therefore unentangled by definition. For this reason, Schmidt rank serves as a measure of entanglement. For more, see, for instance, Nielsen and Chuang \cite{Nielsen:2010vn}.

The SVD performed above implicitly assumes identification of the kets in the case by ordering. For identical particles and an extension for indistinguishable states, a recipe is given by Sciara et al. \cite{Sciara:2017cz}. First, a partial trace is done on the density matrix $\rho$ down to a single particle reduced density matrix $\rho^{(1)}$. For this, a symmetric inner product is performed, following eq. (\ref{sym_inner}). Then, by diagonalizing $\rho^{(1)}$ the obtained eigenstates $|i\rangle$ generate our Schmidt basis $|\tilde{i},\tilde{i}'\rangle$, with the relevant singular value amounting to their contribution being the square root of the eigenvalues. This approach gives a slightly different result than expected by straightforward SVD and becomes more consistent with particle statistics. 

\section{Proof of Schmidt's rank raise}\label{app:SR}
For the Schmidt space projector $S=\sum_k|\lambda_k\rangle\langle\lambda_k|$ defined in the main text, one can revert it to the same basis as an $r$-rank ($r>1$) projector $\Pi$ by reverting Schmidt decomposition. If an SVD approach is used, one has that:
\begin{eqnarray}
S &=& \sum_k|\lambda_k\rangle\langle\lambda_k|\nonumber\\
&=& \sum_{ijk} v_{ik}u_{jk}|ij\rangle\langle ij|v_{ik}^*u_{jk}^* \nonumber\\
&=& \sum_{ijk}|v_{ik}|^2|u_{jk}|^2|ij\rangle\langle ij|.
\end{eqnarray}

If we suppose that only one Schmidt basis exists, the summation over $k$ is trivial, and $|v_i|=|u_j|=1$ in order to have a norm 1 projector. Then, we obtain
\begin{eqnarray}
S - \Pi = \sum_{ij \notin \Pi} |ij\rangle\langle ij|.
\end{eqnarray}
If $S$ contains bases not in $\Pi$, we still have a norm 1 operator,  and a non-entangled state is possible, as assumed. However, if only the bases in $\Pi$ are present, this should be a norm 0 operator, and according to lemma 1, should bear the same rank as $\Pi$. Indeed, if one does not ignore the summation over $k$ and assumes $|v_{ik}|,|u_{jk}|<1$, a finite component of this difference remains, but still one has $|| S - \Pi || < 1$, reassuring that $k=r>1$. Hence, for a state within $\Pi$'s range, Schmidt rank is greater than 1. 

Note that though SVD assumes identification of the states in question, in this case, indistinguishability is introduced ad-hoc with projector $\Pi$, making the above demonstration still valid. One may nonetheless use the transformations discussed for the case of identical bases. In this case, Schmidt decomposition of each of the indistinguishable bases gives the same Schmidt bases and therefore the same projector $S$. When returning from Schmidt bases to computational bases, the underlying ambiguity forces one to take a linear combination much in the same fashion as the one written above, leading to essentially the same calculation. 

\section{Two-qubit maximum correlation}\label{app:maxcorr}

Equation (\ref{q_corr}) in the main text takes for its maximum value
\begin{eqnarray}
\sum_{k}c_k^{(m)}c_k^{(n)},
\end{eqnarray}
when a state lives in the range of the operator $\Pi$. One can simply analyze the possible expansions of operators $A_m$ and $B_n$ according to eq. (\ref{ops}). The simplest case occurs for only two $A$s and $B$s, which must be represented by the same local operators $\mathcal{O}_1(r_i)$ and $\mathcal{O}_2(r_i)$ ($i$ indicating $A$ or $B$)  to bear any correlation. In this case, each $c_k=1$ and $S_n=2$. We may also consider the case of one of the operator alone having two components, but this will not change much since only one of the components will be non-orthogonal to the other operators and generate correlations.

\begin{figure}[tb]
\subfigure[]{\includegraphics[width=0.45\textwidth]{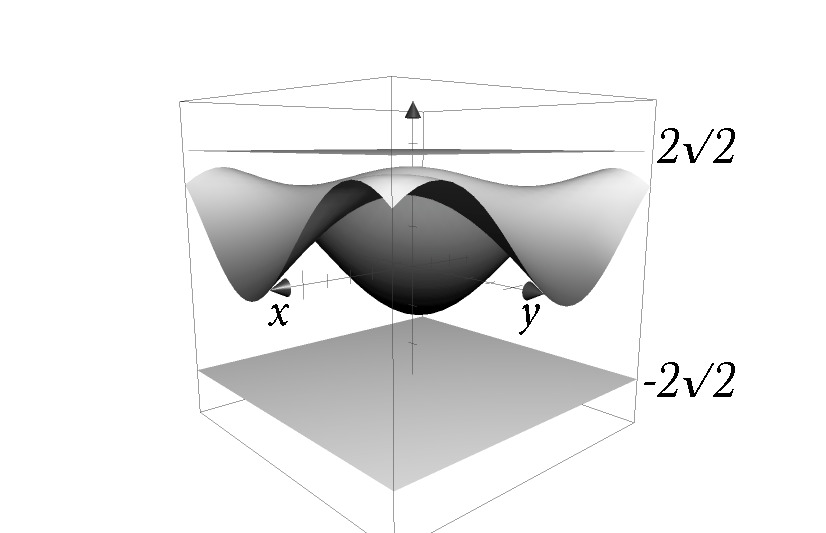}}
\subfigure[]{\includegraphics[width=0.45\textwidth]{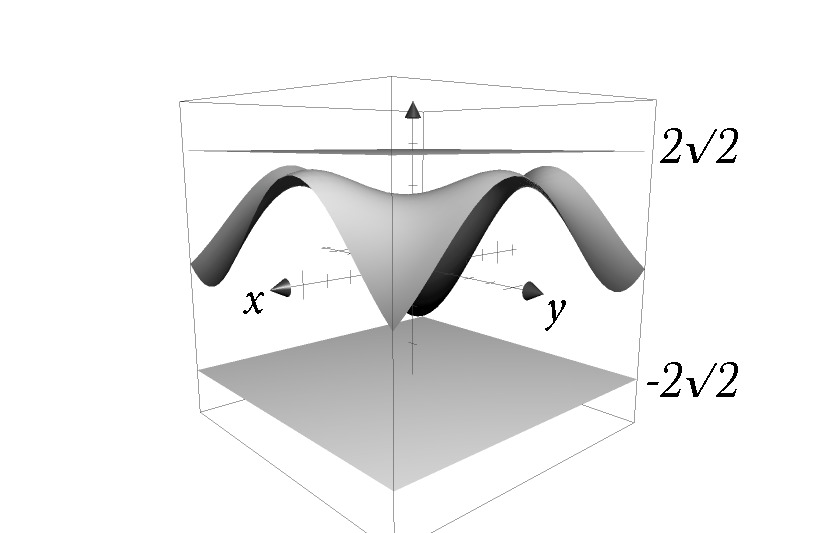}}
\caption{Correlation function in eq. (\ref{corr:1122}) taken (a) $+$, (b) $-$. Given the normalization condition, $c_1^{A(B)}$ is rewritten as $\sin x\,(y)$ and $c_2^{A(B)}$ as $\cos x\,(y)$. The upper and lower plan indicate $z=\pm2\sqrt{2}$, and axes range from $-\pi$ to $\pi$. }
\label{1122}
\end{figure}

We may then consider the case of two operator with two components. First, suppose the case of $A_1=\mathcal{O}_1$, $B_1=\mathcal{O}_1$, $A_2=c^A_i\mathcal{O}_1+c^A_2\mathcal{O}_2$, and $B_2=c^B_i\mathcal{O}_1+c^B_2\mathcal{O}_2$. Their maximum correlations become
\begin{eqnarray}
&\langle A_1B_1\rangle=\pm1,\quad \langle A_1B_2\rangle= \pm c_1^B&\nonumber\\
&\langle A_2B_1\rangle= \pm c_1^A,\quad \langle A_2B_2\rangle= \pm c_1^Ac_1^B\pm c_2^Ac_2^B,\quad&
\end{eqnarray}
which leads to global correlations of the form
\begin{eqnarray}
S_2 = 1 + c_1^A + c_1^B - c_1^Ac_1^B \pm c_2^Ac_2^B. \label{corr:1122}
\end{eqnarray}
By placing the substitutions $c_1^A = \sin x, c_2^A=\cos x, c_1^B=\sin y, c_2^B = \cos y$, we may calculate the highest correlation achieved to be $1+\sqrt{2}$, which can also be verified numerically in by plotting $S_2$ as in fig. (\ref{1122}).

Another two-component possibility lies on the case $A_1=\mathcal{O}_1,A_2=\mathcal{O}_2,B_1=c_1^{(1)}\mathcal{O}_1+c^{(1)}_2\mathcal{O}_2,B_2=c_1^{(2)}\mathcal{O}_1+c_2^{(2)}\mathcal{O}_2$. The correlations for such operators can be calculated as
\begin{eqnarray}
&\langle A_1B_1\rangle=\pm c_1^{(1)},\quad \langle A_1B_2\rangle= \pm c_1^{(2)}&\nonumber\\
&\langle A_2B_1\rangle= \pm c_2^{(1)},\quad \langle A_2B_2\rangle= \pm c_2^{(2)},&
\end{eqnarray}
and
\begin{eqnarray}
S_2= c_1^{(1)} + c_1^{(2)} + c_2^{(1)} - c_2^{(2)}.
\end{eqnarray}
By using the same substitution as in the previous case, it is straightforward to show that $S_n\le 2\sqrt{2}$.
If more components are assumed, the contribution of each component to the total correlation decreases, and the maximum value for $S_2$ decreases together, leaving the maximum value of the known Tsirelson bound.

\section{Nonlocality bound of $n$-body correlation}\label{app:nonlocal}
By rewriting $S_n$ as a function of outcomes equal to 0/1 instead of -1/1, one obtains that
\begin{eqnarray}
\Sigma_n &\le_{\mathrm{H}}& 2^{n-2}\times3 \le_{\mathrm{QT}} (2^{n-2}\times{2+\sqrt{2}}),\\
\Sigma_n &=& \frac{S_n}{2} + 2^{n-1}\\
&=& \sum_{\substack{(x_1,x_2,x_3)\ne (x_1,x_1,x_1) \\ \\ b_1\oplus\cdots\oplus b_n = 0}}p(b_1,\ldots,b_n|x_1,\ldots,x_n)\nonumber\\
&+& \sum_{\substack{(x_1,x_2,x_3) = (x_1,x_1,x_1) \\ \\ b_1\oplus\cdots\oplus b_n = 1}}p(b_1,\ldots,b_n|x_1,\ldots,x_n),\quad
\end{eqnarray}
where $b_i\in\{0,1\}$ are the outputs of the measurement on the $x_i$ basis. By taking two copies of a system, $A$ and $A'$, a global events' probability can be written as
\begin{eqnarray}
%\begin{aligned}
 &&p(b1,\ldots,b_n,b'_1,\ldots,b'_n | x_1, \ldots, x_n, x'_1,\ldots,x'_n)\qquad\nonumber\\
 &&=p(b_1,\ldots,b_n|x_1,\ldots,x_n)\times p(b'_1,\ldots, b'_n|x'_1,\ldots,x'_n).\qquad
%\end{aligned}
%\phantom{3cm}
\end{eqnarray}
Summation of events in $\Sigma_n$ within both copies gives $\Sigma_n^2$. If both copies are not in $\Sigma_n$, their summation add up to $(2^n-\Sigma_n)^2$, where the $2^n$ comes from the $2^n$ different $(x_1,\ldots,x_n)$.

In total, there are $4^n\times4^n/2$ events divided into $4^n$ disjoint sets, each containing $4^n/2$ pairwise exclusive events. One may add to each set a new event $(p,q|A_{ij},A_{\bar{i}\bar{j}})$, with compatible $A_{ij},A_{\bar{i}\bar{j}}$ (i.e. $A_{00},A_{11}$ or $A_{01},A_{10}$) while keeping exclusivity. From E principle, the sum of a set of pairwise exclusive events cannot exceed 1, what translates into an inequality we may call ``E inequality.'' Since there are $4^n$ such set, summing them gives
\begin{eqnarray}
\Sigma_n^2 + (2^n-\Sigma_n)^2 + 4^{n-1} \le 4^n.
\end{eqnarray}
The $4^{n-1}$ term comes form the addition of $(p,q|A_{ij},A_{\bar{i}\bar{j}})$ events that add to unity for a fixed $ij$, existing $4^{n-1}$ of them. From this inequality, one obtains
\begin{eqnarray}
\Sigma_n\le 2^{n-2}\times (2+\sqrt{2}).
\end{eqnarray}
For a detailed derivation, see ref. \onlinecite{Cabello:2015ka}.

\end{document}